\documentclass[11pt]{article}
 \usepackage{amsfonts}
\usepackage{amsmath} 
 \usepackage{amssymb}
 \newfont{\bbbold}{msbm10 scaled \magstep1}

 \def\cF{{\cal F}}
 
 \def\cH{{\cal H}}

 \def\cL{{\cal L}}
 \def\cM{{\cal M}}

 \newfont{\goth}{eufm10 scaled \magstep1}

 \def\a{\alpha}
 \def\b{\beta}
 \def\c{\gamma}
 \def\d{\delta}\def\D{\Delta}
 \def\e{\epsilon}
 
 \def\h{\eta}
 \def\k{\kappa}
 \def\L{\Lambda}

 \def\t{\tau}
 \def\th{\theta}

 \def\ua{\underline{\alpha}}
 \def\ub{\underline{\phantom{\alpha}}\!\!\!\beta}

 \def\una{\underline a}\def\unA{\underline A}
 \def\unb{\underline b}\def\unB{\underline B}
 \def\unC{\underline C}

 \def\unM{\underline M}

 \def\unH{\underline{H}}

 \def\nab{\nabla}

 \def\be{\begin{equation}}\def\ee{\end{equation}}
 \def\bea{\begin{eqnarray}}\def\eea{\end{eqnarray}}
 \def\ba{\begin{array}}\def\ea{\end{array}}

 \newcommand{\hoch}[1]{$^{#1}$}

\begin{document}

\thispagestyle{empty}

 \hfill{KCL-MTH-04-10}

 \hfill{\today}

 \vspace{20pt}

 \begin{center}
 {\Large{\bf Kappa-symmetric Derivative Corrections to D-brane Dynamics}}

 \vspace{30pt}

 {J. M. Drummond\hoch1, S. F. Kerstan\hoch2}

\vspace{15pt}

\begin{itemize}
\item[$^1$] 
Department of Mathematics, Trinity College, Dublin, Ireland.
\item[$^2$]
Department of Mathematics, King's College, London, U.K. 
\end{itemize}

 \vspace{60pt}

 \end{center}

 {\bf Abstract}

We show how the superembedding formalism can be applied to construct 
manifestly kappa-symmetric higher derivative
corrections for the D9-brane. We also show that all correction terms 
appear at even powers of the
fundamental length scale $l$.
We explicitly construct the first potential correction, which 
corresponds to the kappa-symmetric version of the $\partial^4 F^4$,
which one finds from the four-point amplitude of the open superstring.

 {\vfill\leftline{}\vfill \vskip  10pt

 \baselineskip=15pt \pagebreak \setcounter{page}{1}


\section{Introduction}

The derivation of the effective dynamics of branes in string- and M-theory
is a difficult problem. Born-Infeld theory has been shown to describe,
in the limit of slowly varying field strengths, the effective dynamics of
D-branes to all orders in $\a^\prime$ \cite{tseytlin, leigh}. However,
when this  
limit is not valid, ``derivative corrections'' to Born-Infeld theory, 
that is correction 
terms involving the derivatives of the fields, i.e. the world volume field
strength tensor and the transverse location of the brane, must be taken into
account. The first such correction term was calculated in
\cite{at88}. In recent 
years, there has been some progress in the construction of corrections, see for
example \cite{bbg99,wyllard00}. Also, more recently, there have been
several approaches based on supersymmetry. In \cite{cdre02,dkhh03},
for example, the leading 
supersymmetric correction for 
the gauge field in the Abelian case, which is the supersymmetric completion of
$\partial^4 F^4$, has been derived. The full supersymmetric four-point
function was constructed in \cite{dre03}. 
 The leading correction to the
non-Abelian theory, as a description 
of coinciding D-branes, has also been identified using various
methods, see
\cite{ks01,cnt01a,cnt01b,rstz01,grasso,mbm,cdre,dkhh03}. A non-Abelian
generalisation of the four-point function of \cite{dre03} was discussed
in \cite{cm03}. 
However, all these results do not possess the $\k$-symmetry of the
undeformed D-brane actions \cite{aps,cvgnw96}.

The superembedding formalism is a manifestly $\k$-symmetric method for
deriving brane dynamics.  It was developed in \cite{vz,stv,stvz}
  \footnote{ Note that the concept of source and target superspace
appeared already in \cite{gates}. For a review of superembeddings see
\cite{sorokinreview}.}  
and extended into a general framework for the description of
superbranes in \cite{bpstv,bsv,howe96,hsw97,hrs98}. 
Superembeddings have been used to construct higher derivative brane
actions \cite{hl,dhl}.  
For the M2-brane, which has no gauge degrees of freedom, the first
potential $\k$-symmetric 
correction term has been constructed in \cite{hklt03}, using a
(deformed) superembedding. 
In the present work, we use an approach similar to that developed in
\cite{hklt03}. We draw on 
the concept of spinorial cohomologies
\cite{cnt01a,cnt01b,cnt01c,cnt02} to
construct the $\k$-symmetric equations of motion corresponding 
to the $\partial^4 F^4$-term for the D9-brane. We prove that it is the
first such potential derivative correction. 

The dynamics of the D9-brane are captured by two systems of equations
which are  
linked: the torsion equation, which gives the world-volume torsion of the
brane  in terms of the pullback of the target-space torsion, and the
world-volume 
Bianchi 
identities for the gauge field living on the brane. Dimensional analysis puts
a natural constraint 
on the components of the field strength super 2-form, the so-called
$\cF$-constraint. This constraint implies 
Born-Infeld dynamics for the D9-brane. The introduction of a fundamental
length scale allows us to deform the $\cF$-constraint and thereby the Bianchi
identities. 
This leads to derivative corrections in the equations of motion. This approach
is the gauge-field analogue 
of that used in \cite{hklt03}, in which the first potential  derivative
correction for the M2-brane was constructed by deforming 
the constraint on transverse degrees of freedom, the ``embedding constraint''.

Section 2 summarises the superembedding approach as applied to the
D9-brane. This section
essentially contains material covered before
\cite{bandits,bppst01,dh01,kerstan02}, 
but serves to prepare the ground for the discussion of derivative
corrections in this approach.

In section 3 we prove generally that deformations of D9-dynamics and
also D=10 SYM (or, in fact D=10 Yang-Mills coupled to a spinor of
canonical dimension), can only exist with coefficients of even powers of the
fundamental length scale $l$, or, if one likes, integer powers of
$\a'$. 
We then explicitly deform the standard constraints on the
D9-brane in powers of the fundamental length scale $l$. 
We show that the first potential deformation allowed, which is
at $l^2$, does not exist. We then explicitly construct 
the deformation at the next order, $l^4$, cubic in fields. As in the case of
the M2-brane, the problem turns out to be a spinorial 
cohomology problem as described in \cite{cnt01a,cnt01b}. We find a
unique solution cubic in fields (but have not checked 
consistency at higher orders in fields or higher powers of l). 

In section 4 we show the effect of the deformed constraint on the
kappa-symmetric equations of motion for the worldvolume theory.

We conclude with some comments on higher order terms and other branes.

\section{The D9-brane as a superembedding}

Superembeddings are the generalisation of surface theory to
supermanifolds. They are well suited for describing gauge invariant
brane dynamics for the branes which arise in superstring theories and
M-theory. The fermionic gauge symmetry, kappa-symmetry, which is present in
the Green-Schwarz description of such objects has a natural geometrical
interpretation as the odd part of the local reparametrisation invariance of
the embedded manifold (the brane). The parameter of the kappa
symmetry in the Green-Schwarz formalism is replaced by an odd vector field on
the worldvolume of the brane. 

We consider such an embedding, with $\cM$ labelling the worldvolume of a
p-brane and $\underline{\cM}$ labelling the $D$ dimensional target space,
which we take to be flat. The embedding is a map, 

\be
f : \cM \longrightarrow \underline{\cM}.
\ee

The cotangent frame on the target space, $E^{\unA} = (E^{\una},E^{\ua})$ can be
pulled back to the worldvolume via the pullback map, $f^*$, and expressed in
terms of the cotangent frame on the worldvolume, $E^A = (E^a,E^{\a})$, 

\be
f^* E^{\unA} = E^A E_{A}{}^{\unA}. 
\ee

We use the convention that Latin indices refer to the bosonic directions and
Greek indices to the fermionic directions. Capital letters are used for both,
$A=(a,\a)$. Underlined indices refer to target space quantities while those
without an underline refer to the worldvolume. Target space indices are split
by the embedding into directions tangential and normal to the
worldvolume. Normal directions are denoted with a primed index, $a'$ or
$\a'$. The matrix $E_{A}{}^{\unA}$ is called the embedding matrix and contains
the geometrical information of the embedding. It can be put into a general
form,  

\be
E_{A}{}^{\unA} =
\left(
\ba{rr}
 u_{a}{}^{\una} & \L_{a}{}^{\b'}u_{\b'}{}^{\ua} \\
 \Psi_{\a}{}^{b'}u_{b'}{}^{\una} & u_{\a}{}^{\ua} +
h_{\a}{}^{\b'}u_{\b'}{}^{\ua} 
\ea
\right).
\ee \label{embeddingmatrix}

Here the matrix $u_{\una}{}^{\unb}$ is an element of $SO(1,D-1)$ and
$u_{\ua}{}^{\ub}$ is the corresponding element of $Spin(1,D-1)$. For
certain superembeddings, the standard constraint

\be
E_{\a}{}^{\una}=0, 
\ee

implies the equations of motion for the worldvolume
supermultiplet. Embeddings describing the M2-brane and M5-brane are
examples of this. For some superembeddings of low bosonic codimension,
a second constraint, called the $\cF$-constraint, is required. In the
D-brane superembeddings there is a closed super three-form, $\unH$, in
the type II supergravity background. One introduces a worldvolume
two-form, $\cF$, which satisfies $d\cF = f^*\unH$. The standard
$\cF$-constraint takes the form

\be
\cF_{\a \b} = \cF_{\a b} = 0.
\ee

In the absence of an explicit length scale, with the standard
embedding condition, no objects of negative mass dimension
appear. Dimensional analysis then forces the above form of the 
$\cF$-constraint. The constraint implies that the multiplet on the
worldvolume of the D-brane is given by the Maxwell supermultiplet
(which is on-shell) and thus the fields are constrained to satisfy
their equations of motion. One can write the two-form, $\cF$, as 

\be
\cF = F + f^*\unB.  
\ee

Here, $\unB$, is the two-form potential for $\unH$ in the background,
$d\unB = \unH$, and $F$ is a two-form field strength for the
worldvolume one-form gauge potential, $A$, satisfying $dA = F$. In a
flat background, one can give an explicit solution for $\unB$ in terms
of the target space coordinates. The deformation of
the standard Maxwell constraints on the components of $F$, which defines
supersymmetric (Born-Infeld) theory living on the worldvolume of the
D-brane, can then readily be deduced from the $\cF$-constraint. 
This was carried out explicitly for the D9-brane of IIB in
\cite{kerstan02}.    

In \cite{hklt03} a
deformation of the embedding constraint was used to describe higher
derivative corrections to the M2-brane equations of motion. Here we will be
constructing derivative deformations of D9-brane dynamics.
In the case of the D9-brane (and other space-filling branes), the 
embedding condition is satisfied without loss of generality ( so
$\Psi$ in (\ref{embeddingmatrix}) is 0) 
since there are no normal bosonic directions. Thus the embedding
matrix must remain 
undeformed even when derivative corrections are to be
included. Instead, one must deform the $\cF$-constraint.  
Note that for space filling branes, we can also always take the matrix
$u$ in (\ref{embeddingmatrix}) to 
be the identity without loss of generality, thus identifying the
bosonic cotangent frame of the two  
supermanifolds. We therefore choose for our D9-brane embedding

\be
E_{A}{}^{\unA} =
\left(
\ba{rr}
 \d_{a}{}^{\una} & \L_{a}{}^{\b'}\d_{\b'}{}^{\ua} \\
 0& \d_{\a}{}^{\ua} +
h_{\a}{}^{\b'}\d_{\b'}{}^{\ua} 
\ea
\right).
\ee \label{embeddingmatrix2}

Here, and from now on, the Greek indices $\a$ are 16 component
Majorana-Weyl spinor indices of $Spin(1,9)$. The two chiralities are
denoted by upstairs and downstairs indices. The target space
indices $\ua$ can be split into the pair $\a i$ where the index
$i$ is an $SO(2)$ index.

The main aim of this paper is to describe how one can systematically
deform this constraint to include derivative corrections to Born-Infeld
theory. The deformations are described in terms of spinorial cohomology and
will be discussed in the next section. 

We now focus on the supergeometry of the D9-brane superembedding and
show how one can deduce the equations of motion given the
standard $\cF$-constraint . We consider the embedding of
$N=(1,0)$, $D=10$ superspace into flat $N=(2,0)$, $D=10$ superspace.   
The worldvolume geometry is constrained by the torsion equation which gives
the worldvolume torsion in terms of the target space torsion.
We have

\be
T^A E_{A}{}^{\unA} = dE^A E_{A}{}^{\unA} = f^* d E^{\unA} = f^* T^{\unA}.
\ee

The target space will be assumed flat throughout and hence the only non-zero
component of the target space torsion is at dimension zero,

\be
T_{\a i \b j}{}^{c} = -i \d_{ij} (\c^c)_{\a \b}.
\ee

In components the torsion equation reads,

\be
\nab_A E_{B}{}^{\unC} - (-1)^{AB}\nab_B E_{A}{}^{\unC} +
T_{AB}{}^{C}E_{C}{}^{\unC} 
= (-1)^{A(B+\unB)}E_{B}{}^{\unB}E_{A}{}^{\unA} T_{\unA \unB}{}^{\unC}.
\ee

Analysing this equation level by level in dimension and substituting
(\ref{embeddingmatrix2}), we find: 

{\bf dimension 0 :}
\be
T_{\a \b}{}^{c} = -i(\d_{\a}^{\c} \d_{\b}^{\c} + h_{\a}{}^{\c}
h_{\b}{}^{\d}) (\c^c)_{\c \d} 
\ee

{\bf dimension $\tfrac{\bf 1}{\bf 2}$ :}
\begin{align}
T_{\a \b}{}^{\c} &= 0, \\
T_{\a b}{}^{c} &= i \L_{b}{}^{\b} h_{\a}{}^{\c} (\c^c)_{\c \b}, \\
2\nab_{(\a} h_{\b )}{}^{\c} &= - T_{\a \b}{}^{a} \L_{a}{}^{\c}.
\end{align}

{\bf dimension 1 :}
\begin{align}
T_{\a b}{}^{\c} &= 0, \\
T_{ab}{}^{c} &= -i \L_{b}{}^{\b} \L_{a}{}^{\a}(\c^c)_{\a \b}, \\
\nab_{\a} \L_{b}{}^{\c} - \nab_b h_{\a}{}^{\c} &
= -i \L_{b}{}^{\b} \L_{c}{}^{\c} h_{\a}{}^{\d} (\c^c)_{\b \d} 
\end{align}

{\bf dimension $\tfrac{\bf 3}{\bf 2}$ :}
\begin{align}
T_{ab}{}^{\c} &= 0, \\
2\nab_{[a}\L_{b]}{}^{\c} &= i\L_{b}{}^{\b} \L_{a}{}^{\a} \L_{c}{}^{\c}
(\c^c)_{\a \b}. 
\end{align}

Next we proceed to the Bianchi identity for the two-form, $\cF$,

\be
d\cF = f^*\unH.
\ee

The non-zero components of the three-form, $\unH$, in a flat IIB background,
are given by 

\be
H_{\a i \b j c} = -i (\c_c)_{\a \b} H_{ij}. 
\ee

There are two linearly independent closed forms, given by 

\be
H_{ij} = (\t^1)_{ij} \text{ and } H_{ij} = (\t^3)_{ij},
\ee

with $\t^i$ being the Pauli matrices. We choose the first of these solutions,
$H_{ij} = (\t^1)_{ij}$.

In components, the Bianchi identity reads

\be
3\nab_{[A} \cF_{BC]} + 3T_{[AB}{}^{D}\cF_{|D|C]} =
(-1)^{A(B+\unB+C+\unC)+B(C+\unC)} E_{C}{}^{\unC} E_{B}{}^{\unB} E_{A}{}^{\unA}
H_{\unA \unB \unC}. 
\ee

Again, we analyse this level by level in dimension. 

{\bf dimension $-\tfrac{\bf 1}{\bf 2}$ :}
\be
\nab_{(\a}\cF_{\b \c)} + T_{(\a \b}{}^{d}\cF_{d \c)} = 0.
\ee \label{-12bi}

{\bf dimension 0 :}
\be
2\nab_{(\a} \cF_{\b)c} + \nab_c\cF_{\a \b} + T_{\a \b}{}^{d}\cF_{dc} + 
2T_{(\a c}{}^{d}\cF_{d \b)} = E_{\a}{}^{\ua} E_{\b}{}^{\ub} H_{\ua \ub c}.
\ee

{\bf dimension $\tfrac{\bf 1}{\bf 2}$ :}
\be
\nab_{\a}\cF_{bc} -2\nab_{[b}\cF_{\a c]} + 2T_{\a [b}{}^{d}\cF_{|d|c]} +
T_{bc}{}^{d} \cF_{d \a} = -2E_{[b}{}^{\ub}E_{\a}{}^{\ua} H_{\ua \ub c]}.
\ee

{\bf dimension 1 :}
\be
\nab_{[a} \cF_{bc]} + T_{[ab}{}^{d} \cF_{|d|c]} = E_{[b}{}^{\ub}
  E_{a}{}^{\ua} H_{\ua \ub c]}
\ee

If we impose the standard constraints on $\cF$ given by

\be
\cF_{\a \b} = \cF_{\a b} = 0
\ee

then the Bianchi identity at dimension $-\frac{1}{2}$ is automatically
satisfied. We will deform the $\cF$-constraint in the next section but
for now we work with the undeformed constraint.

If we linearise the above system of equations we find that the
solution can be written quite simply. We can split $\L$ into its two
irreducible representations, 

\be
\L_{a}{}^{\a} = (\c_a)^{\a \b} \psi_\b + \hat{\L}_{a}{}^{\a},
\ee

where $\hat{\L}$ is gamma-traceless. Then, we substitute our results
from the torsion equation into the Bianchi identity. From the dimension
0 component we find, by contracting with $\left(\c^1\right)^{\a\b}$
and $\left(\c^5\right)^{\a\b}$ 
\be
h_{\a}{}^{\b} = \frac{1}{4} \cF_{cd}(\c^{cd})_{\a}{}^{\b}.
\ee
At dimension $\frac{1}{2}$ we find 
\bea
\psi_{\d} &=& 0, \label{fermioneom} \\
\nab_{\a} \cF_{bc} &=& 2i \hat{\L}_{[b}{}^{\b} (\c_{c]})_{\a \b}
\eea
From the dimension 1 Bianchi identity we get
\bea
\nab_{\a} \hat{\L}_{b}{}^{\b} &=& \frac{1}{4} \nab_b \cF_{cd}
(\c^{cd})_{\a}{}^{\b}\\ 
\nab_{[a} \cF_{bc]} &=& 0 \\
\nab^a \cF_{ab} &= 0 \label{bosoneom}
\eea
The dimension $\frac{3}{2}$ Torsion equation contains
\bea
\nab_{[a} \hat{\L}_{b]}{}^{\a} &=& 0 \\
\nab^a \hat{\L}_{a}{}^{\a} &=& 0
\eea

The relation (\ref{fermioneom}) is the linearised fermionic equation
of motion for the worldvolume theory.  One can see this by going to
static gauge, which is defined by the identification of the
worldvolume coordinates $x^a$ and $\th^\a$ with the target space
coordinates $x^{\una}$ and $\th^{\a 1}$. The second spinorial
coordinate becomes the field on the worldvolume, $\th^{\a 2} =
\zeta^\a (x^a, \th^\a)$, which is (at lowest order) the spinor field
strength superfield of the deformed Maxwell theory. In this gauge,
$\L_{a}{}^{\a} = \nab_a \zeta^\a$, so, at the linear level, the above
equation is the Dirac equation for $\zeta$. The full non-linear
equation is the supersymmetric Born-Infeld equivalent. The relation
(\ref{bosoneom}) is the linearised bosonic equation of motion for the
for the worldvolume theory. This obviously has the form of the Maxwell
equation for the vector field $A_a$.

The linear relations constrain the fields $\cF_{ab}$ and
$\hat{\L}_{a}{}^{\a}$ and their derivatives to lie in certain
representations of $Spin(1,9)$ (see the appendix for notation)

\begin{align}
&\nab_{a_1}...\nab_{a_n}\cF_{bc} \text{ in the irrep } (n1000),
  \label{rep1} \\
&\nab_{a_1}...\nab_{a_{n-1}}\L_{a_n}{}^{\a} \text{ in the irrep } (n0010).
\label{rep2}
\end{align}

The non-linear corrections to the linearised equations can then be regarded as
an expansion in the above objects (i.e. the fields satisfying their lowest
order relations). This is also how we will treat the additional non-linear
corrections which appear when we deform the $\cF$-constraint.

The non-linear corrections to the undeformed theory can be derived from the
system of torsion equations and Bianchi identities. We can decompose the
matrix, $h_{\a}{}^{\b}$, into irreducible representations,

\be
h_{\a}{}^{\b} = h \d_{\a}^{\b} + h_{ab} (\c^{ab})_{\a}{}^{\b} 
+ h_{abcd} (\c^{abcd})_{\a}{}^{\b}.
\ee

As we have seen, at the linearised level, only the two-form component
of $h_{\a}{}^{\b}$ is non-zero and that this is the two-form component
of $\cF$, up to a factor. 
The non-linear parts of $h_{\a}{}^{\b}$ will be denoted by
$h'_{\a}{}^{\b}$ and up to cubic order in $\cF$ we find 

\begin{align}
h' &= 0, \\
h'_{ab} &= \frac{1}{8}(\cF_{ac}\cF^{cd}\cF_{db} +
\frac{1}{4} \cF_{cd}\cF^{cd}\cF_{ab}), \\
h'_{abcd} &= \frac{1}{16.2.6.4!} \e_{abcdefghij}\cF^{ef}\cF^{gh}\cF^{ij}. 
\end{align}

The equations at dimension $\tfrac{1}{2}$ imply, 

\begin{align}
\nab_\a \cF_{bc} = &2i \L_{[b}{}^{\b} (\c_{c]})_{\a \b} 
-2i (\c^d)_{\c \b} h_{\a}{}^{\c} \cF_{d[c}\L_{b]}{}^{\b}.
\end{align}

This relation is the supervariation of $\cF$ in the non-linear
theory. Note that the first term contains a linear piece and also a
non-linear piece (the gamma-trace), given by

\begin{align}
\psi_\d = \frac{1}{10} (\c^a)_{\a \d} \L_{a}{}^{\a} = 
\frac{1}{700}[&-6(\c^d)_{\b \c} h_{\a}{}^{\c} \cF_{dc} \L_{b}{}^{\b}
  (\c^{bc})^{\a}{}_{\d} \notag \\
&-2i (\c_a)^{\a \b} \nab_\a h'_{\b}{}^{\c} (\c^a)_{\c
    \d} \notag \\
&- (\c_a)^{\a \b} h_{\a}{}^{\e} h_{\b}{}^{\h} (\c^c)_{\e \h}
  \L_{c}{}^{\c} (\c^a)_{\c \d}]
\end{align}

This is the non-linear fermionic equation of motion for the
worldvolume theory. One can use this equation to determine $\psi$
order by order in fields once $h_{\a}{}^{\b}$ is known.

At dimension one we have

\begin{align}
\nab_{[a} \cF_{bc]} = &-i\L_{[b}{}^{\b} \L_{a}{}^{\a}
  (\c^d)_{\a \b} \cF_{c]d} \\
\nab^b \cF_{ab} = &\frac{1}{8}[10(\c_a)^{\a \e} \nab_\a \psi_\e  
- \nab_b h'_{\a}{}^{\c} (\c^b \c_{a})_{\c}{}^{\a} +i\L_{b}{}^{\b}
  \L_{c}{}^{\c} h_{\a}{}^{\d} (\c^c)_{\b \d} (\c^b \c_a)_{\c}{}^{\a}], 
\end{align}

which are, respectively, the non-linear Bianchi identity for
$\cF_{ab}$ and the non-linear bosonic equation of motion.
We also find the non-linear supervariation of $\cF_{ab}$,

\be
\nab_{\a} \L_{a}{}^{\b} =  \nab_b h_{\a}{}^{\c} - i\L_{b}{}^{\b}
\L_{c}{}^{\c} h_{\a}{}^{\d} (\c^c)_{\b \d}. 
\ee

These equations determine everything as a series in the the fields of
the linearised theory, i.e. $\cF_{ab}$ and $\hat{\L}_{a}{}^{\a}$ and
their derivatives, in the representations (\ref{rep1},\ref{rep2}). 
Thus we have seen that the $\cF$-constraint gives the full non-linear
dynamics of the D9-brane worldvolume theory.

Kappa symmetry in this formulation is manifest. The equations we have
written down are invariant under diffeomorphisms of the
worldvolume. The odd diffeomorphisms are precisely the kappa symmetry
transformations \cite{stv,bpstv,bsv,hrs98}.

\section{Derivative corrections}

Our aim is to understand how to introduce a deformation into the theory which
will allow a manifestly kappa-symmetric treatment of higher derivative
terms. We will follow the analysis presented in \cite{hklt03}, where
derivative corrections to the worldvolume theory of the eleven-dimensional
supermembrane are discussed in the superembeddings framework. In the
supermembrane setting, such corrections can be introduced via a deformation of
the embedding condition $E_{\a}{}^{\una} = 0$. 
In contrast, the only freedom we have here is to relax the standard constraint
on the worldvolume 2-form $\cF$. Thus we have to allow the components
$\cF_{\a \b}$ (dimension $-1$) and $\cF_{\a b}$ (dimension
$-\tfrac{1}{2}$) to be given in terms of the covariant degrees
of freedom $\cF_{ab}$ (dimension $0$) and $\L_{a}{}^{\a}$ (dimension
$\tfrac{1}{2}$), and their derivatives. We must
introduce a parameter, $l$, with unit negative dimension in order to respect
the negative dimensionality of the components of $\cF$. We then search for
possible deformations of $\cF_{\a \b}$ and $\cF_{\a b}$ order by order in $l$. 

In addition to the constraints described in the previous section, the
deformations are constrained by the Torsion equation and Bianchi identity
which must still be satisfied. Furthermore, there is a degree of redundancy in
these quantities which can be absorbed by field redefinitions, of which there
are two types. Firstly, we can redefine the embedding coordinates

\be
z^{\unM} \longrightarrow z^{\unM} + (\d z)^{\unM}.
\ee
  
This transformation is, equivalently, a target space diffeomorphism. The
second type of redefinition is a shift of the one-form potential

\be
A \longrightarrow A + \d A.
\ee

These redefinitions can be used to remove the gamma-trace part of
$\cF_{\a b}$. The quickest way to see this is as follows \footnote{We
  thank Paul Howe for this neat argument.}. When we perform a
diffeomorphism of the target space by a vector field $v$, we see that
the three-form $\unH$, changes by  

\be
\cL_v \unH = (i_v d + d i_v) \unH = d i_v \unH.
\ee

The second equality follows from the closure of $\unH$. This can be
thought of as a change in the two-form potential $\unB$, by $i_v
\unH$. Examining the pullback of $\unB$ we see that there is a change (at
lowest order) to the quantity $\cF_{\a b}$

\be
\cF_{\a b} \longrightarrow \cF_{\a b} + (\c_b)_{\a \b} v^{\b 2},
\ee

if we choose $v^{\una} = v^{\a 1} = 0$. Therefore such field redefinitions can
be used to remove the gamma-trace part of $\cF_{\a b}$.

The second type of field redefinition is of the same form as one has in the
problem of deforming $N=1$, $D=10$ Yang-Mills theory \cite{cnt01a}. 
There are two parts to the shift in the one-form potential, the vector part
$(\d A)_a$, and the spinor part $(\d A)_\a$. In the Yang-Mills case
the vector shift can be used to remove the vector part of $F_{\a \b}$,
leaving only the anti-self-dual 5-form part. Similarly, we can use it
here to remove the vector part of $\cF_{\a \b}$ and so the
deformations we are looking for are also given by anti-self-dual
5-forms $J$:

\be
\cF_{\a\b} = J^{abcde} \left(\c_{abcde}\right)_{\a\b}
\ee

In \cite{cnt01a}, it was noted that the Bianchi identity together with
the existence of the spinorial field redefinitions $(\d A)_\a$,
implied that each deformation can be identified with an element of a
particular spinorial cohomology. This can be seen as follows.

The relevant sequence is one of irreducible representations of
$Spin(1,9)$

\be
(00000) \phantom{a}^{\D_0} \!\!\!\!\!\!\!\!\!\! \longrightarrow
(00001) \phantom{a}^{\D_1} \!\!\!\!\!\!\!\!\!\! \longrightarrow
(00002) \phantom{a}^{\D_2} \!\!\!\!\!\!\!\!\!\! \longrightarrow
(00003) \phantom{a}^{\D_3} 
\!\!\!\!\!\!\!\!\!\! \longrightarrow
... \longrightarrow
(0000n) \phantom{a}^{\D_n} \!\!\!\!\!\!\!\!\!\! \longrightarrow
... \label{seq} 
\ee

The irreps are respectively a scalar, a downstairs spinor, an anti-self-dual
5-form, an anti-self-dual 5-form spinor, etc. The operations, $\D_n$, are
given by the action of a spinorial derivative followed by projection onto the
irrep $(0000 \hspace{2pt} n+1)$. It follows from the algebra of spinorial
derivatives that there is a nilpotence condition,
given by

\be
\D_{n+1} \D_n = 0.
\ee

In the case of the D9-brane, the algebra of spinorial derivatives is given by
the worldvolume torsion. At lowest order in fields this is the standard
torsion and so one can use the same arguments as in the $N=1$, $D=10$
Yang-Mills case \cite{cnt01a} to classify the deformations.

The nilpotence condition allows the definition of the cohomology of $\D_n$ by 

\be
\cH^n = \frac{ {\rm Ker} \D_n}{{\rm Im} \D_{n-1}} .
\ee
 
We find from the dimension $-\frac{1}{2}$ Bianchi identity \ref{-12bi}
that the anti-self-dual five-form,  
$\cF_{\a \b}$, must satisfy the condition,

\be
\nab_{\c} J_{abcde} - (\c_{f[a})_{\c}{}^{\a} \nab_{\a} J_{bcde]}{}^{f}
-\frac{1}{2} (\c_{fg[ab})_{\c}{}^{\a} \nab_{\a} J_{cde]}{}^{fg} = 0.
\ee

This condition is the statement that the $(00003)$ representation contained in
the quantity $\nab_{\a} J_{abcde}$ vanishes, i.e. 

\be
J \in {\rm Ker} \D_2.
\ee

We call this the closure condition. There are some $J$ for which the closure
condition is satisfied trivially. Such $J$ correspond to redefinitions, $\d
A$, of the spinorial part of the gauge potential 1-form and as such do not
correspond to genuine deformations. They are trivially closed because they are
given by $J = \D_1 \d A$ and closure follows from the nilpotence
condition. Such $J$ are referred to as exact. It therefore follows that the
genuine deformations are given elements of the cohomology $\cH^2$. We
will explicitly calculate such objects. 

We will be considering an expansion in numbers of fields. Our aim is calculate
the first allowed non-zero $J$ which is a genuine perturbation of the
Born-Infeld theory. The objects available to construct $J$ are the same
degrees of freedom present in the linearised theory and they can be taken to
satisfy their linearised equations of motion and supersymmetry transformation
rules since these are true up to higher orders in fields. We therefore have
$\L_{a}{}^{\a}$ and the field strength tensor $\cF_{ab}$ and their
derivatives, in the representations constrained by the linearised
theory (\ref{rep1}, \ref{rep2}).


\subsection{All order constraints from dimensional analysis}

We have identified deformations of the Born-Infeld theory with
anti-self-dual 5-forms, $J$, of mass dimension $-1$.
The mass dimensions of the quantities from which we can build
such $J$ are: 
\bea
[l] &=& -1\\
\left[\partial_a \right] &=& 1\\
\left[\hat{\L}_a{}^{\a}\right] &=& \frac{1}{2}\\
\left[\cF_{ab}\right] &=& 0.
\eea
The only quantity that comes with a non-integer mass dimension is the
fermion $\L$ (mass dim. $\frac{1}{2}$). 
A deformation of $\cF_{\a\b}$ (mass dim. 1) of fractional mass
dimension would therefore have to contain an odd 
number of $\L$s. However, $\cF_{\a\b}$ is a boson, and hence must
contain an even number of fermions. 
Therefore, all deformations come at integer powers in mass dimension.

Any deformation can be written, schematically, as
\be
\cF_{\a\b} = J^{abcde} \left(\c_{abcde}\right)_{\a\b} = l^x
\partial^{k} \L^{2n} \cF^m \label{defs} 
\ee
Apart from derivatives and the fields $\L$ and $\cF$, the expression
can also contain the volume form $\e^{10}$ and the metric
$\eta$ which carry an even number of vector indices. Dimensional
analysis of the above equation implies 
\be
-1 = -x + k + n
\ee
For $x$ odd, this implies that $k+n$ is even. Each pair of spinor
indices can be replaced with an odd number (which we call $p_i$ for
the $i$th pair) of vector indices by
contraction with $\c^a$, $\c^{abc}$ or $\c^{abcde}$. The number of
vector indices is then
\be
k + 2n + \sum_{i=1}^{n} p_i + 2m  = (k + n) + \sum_{i=1}^{n}
(p_i + 1) + 2m = \text{even}, 
\ee
and hence there is no way to construct a 5-form, which has an odd
number of vector indices.

Essentially the same argument also goes through for any deformations
of the (Abelian or non-Abelian) $N=1$ Super Yang-Mills Lagrangian in
$D=10$ (or indeed Yang-Mills coupled to a spinor of canonical
dimension). In this case the deformation is a scalar constructed from
the fields $W^\a$ (dimension $-\tfrac{1}{2}$) and $F_{ab}$ (dimension $0$)
and their covariant derivatives. Again one finds only even integer
powers of $l$ (or integer powers of $\a'$) in the deformations. Note
that this is not the case in the case of the deformed M2-brane
\cite{hklt03} where one has to check explicitly that there are no
deformations at $l^3$, for example.

\subsection{Perturbative construction of deformations}

To construct $J$ we need to find a bosonic anti-self-dual 5-form (rep (00002))
inside 
the tensor product of some number of the representations
(\ref{rep1}, \ref{rep2}). There are none 
linear in the above objects (obviously) and nor are there any
quadratic. Therefore we consider terms cubic in the fields. The first such
objects are at order $l^2$. There are three of the form $\L \L \cF$ and one of
the form $\nab \cF \cF \cF$:

\begin{align}
J_1 &= l^2\L_{[a_1}^{\a} \L_{a_2}{}^{\b} \cF_{a_3 a_4} (\c_{a_5]})_{\a \b}, \\
J_2 &= l^2\L_{b}{}^{\a} \L^{b \b} \cF_{[a_1 a_2} (\c_{a_3 a_4
    a_5]})_{\a \b}, \\ 
J_3 &= l^2\L_{b}{}^{\a} \L_{[a_1}{}^{\b} \cF^{b}{}_{a_2} (\c_{a_3 a_4
    a_5]})_{\a \b}, \\ 
J_4 &= l^2\nab_b \cF_{[a_1 a_2} \cF^{b}{}_{a_3} \cF_{a_4 a_5]}, 
\end{align}

where anti-self-duality is implicit. To see which of these are exact, we must
consider all possible spinors $(\d A)_\a$ 
which are cubic in the fields at order $l^2$. There are two:

\begin{align}
\d A_1 &= l^2\L_{a}{}^{\b} \cF^{ab} \cF_{bc} (\c^c)_{\a \b}, \\
\d A_2 &= l^2\L_{a}{}^{\b} \cF^{ab} \cF^{cd} (\c_{bcd})_{\a \b}.
\end{align}

To calculate $\D_1 \d A$,
we form the quantity $(\c_{a_1...a_5})^{\a \b} \nab_\a \d A_\b$ for both
of the above. The results are combinations of the $J$s above. We find that

\begin{align}
\D_1 \d A_1 &= J_2, \\
\D_1 \d A_2 &= J_1 -\frac{1}{4} J_4. \\
\end{align}
 
So the above combinations of $J$s are exact. To check whether the remaining
two linearly independent combinations are closed (i.e. $\D_2 J = 0$) we need
to consider the possible anti-self-dual 5-form spinors, $C$, (rep (00003))
which are cubic in the fields at order $l^2$. There are two of these, given by
the (00003) parts of:

\begin{align}
C_1 &= l^2\L_{[a_1}{}^{\a} \L_{a_1}{}^{\b} \L_{a_3}{}^{\c} 
(\c_{a_4 a_5] b})_{\a \b} (\c^{b})_{\c \d}, \\
C_2 &= l^2\L_{[a_1}{}^{\a} \nab^b \cF_{a_2 a_3} \cF_{a_4 a_5]} (\c_b)_{\a \b}.
\end{align}

Applying a spinor derivative to the remaining $J$s we find, 

\begin{align}
\D_2 J_3 &= -i C_1 \\
\D_2 J_4 &= i C_2 .
\end{align}

Hence the only closed combinations of $J$s are exact and we find the
cohomology is trivial at this order in $l$.  

We therefore proceed to higher order in $l$. The next possibility is $l^4$ and
we find nineteen possible anti-self-dual 5-forms cubic in the fields. These
are listed in the appendix. There are fourteen spinors $(\d A)_\a$ cubic in 
the fields at order $l^4$, however, there are at most twelve independent field
redefinitions because there are two combinations which can be written in the
form, 
 
\be
(\d A)_\a = \nab_\a \phi,
\ee

for some scalar $\phi$. Such spinors are exact in the sequence (\ref{seq}) and
hence they give no field redefinition. By direct calculation one can verify
that indeed the remaining twelve are independent.
There are six possibilities in the
representation (00003). By counting one can see that there is at least one $J$
which is closed since there are 19 $J$s in total, twelve of which are field
redefinitions, leaving 7 possibilities. There are 6 constraints from the
closure condition so there is at least one solution. Again a direct
calculation reveals that there is indeed only one solution, which can be
written in the form,

\be
J_{a_1 a_2 a_3 a_4 a_5} = l^4\nab_b \L_{c}{}^{\a} \L^{b \b} \nab^c
\cF_{[a_1 a_2} (\c_{a_3 a_4 a_5]})_{\a \b} - \text{ dual }.
\ee

This is our main result. The above term represents the first
supersymmetric and kappa-symmetric deformation of the Born-Infeld
theory of the D9-brane at lowest order in fields and lowest order in
the dimensionful parameter $l$.
The details of the derivation are given in the appendix, where all the
relevant quantities in the representations (00000), (00001), (00002)
and (00003) are written explicitly.

\section{Derivative corrections to the equations of motion}

The derivative corrections to the equations of motion can be simply
computed by including a non-zero $\cF_{\a \b}$ in the
Bianchi identity. We work up to order three in fields since 
$\cF_{\a \b}$ is computed up to this order. The corrections to the
various quantities given in section 2 are denoted below by a
superscript (3,1), referring to order 3 in fields and order 1 in the
deformation $J$.

The Bianchi identity at dimension $-\tfrac{1}{2}$, gives the gamma
traceless part of $\cF_{\a b}$ in terms of $\cF_{\a \b}$. We find

\be
\hat{\cF}_{\a b} = \frac{i}{10} (\c_b)^{\b \c} \nab_\c \cF_{\a \b}.
\ee

At dimension zero we find the leading corrections to the different
irreps in $h_{\a}{}^{\b}$.

\begin{align}
h^{(3,1)} &= 0, \label{hJscalar} \\
h^{(3,1)}_{ab} &= \frac{1}{8} (\c^{cde})^{\a \b} \nab_\a \nab_\b
J_{abcde}, 
\label{hJ2form} \\
h^{(3,1)}_{abcd} &= -\frac{1}{24}(\c_{[a}{}^{ef})^{\a \b} \nab_\a \nab_\b
    J_{bcd]ef} -\frac{7i}{6} \nab^e J_{abcde}. \label{hJ4form}
\end{align}  
   
At dimension $\tfrac{1}{2}$ we have the correction to the
supervariation of $\cF$,

\be
(\nab_\a \cF_{bc})^{(3,1)} = 2i(\c_{bc})^{\c}{}_{\a} \psi^{(3,1)}_\c +
\frac{2i}{10} (\c_{[c})^{\c \d} \nab_{b]} \nab_\c \cF_{\a \d} ,
\ee

and the correction to the fermionic equation of motion,

\be
\psi^{(3,1)}_{\a} = \frac{6}{10.700} \nab_b \nab_\e \cF_{\b \d}
(\c_c)^{\e \d} (\c^{bc})^{\b}{}_{\a} 
-\frac{2i}{700} (\c^h)^{\e \b} \nab_\e
h^{(3,1)}{}_{\b}{}^{\c}(\c_h)_{\c \a}.
\ee

At dimension one we have no correction to the component Bianchi
identity for $\cF_{ab}$,

\be
(\nab_{[a} \cF_{bc]})^{(3,1)} = 0,
\ee

and the correction to the bosonic equation of motion,

\be
(\nab^a \cF_{ab})^{(3,1)} = -\frac{10}{8} (\c_a)^{\a \e} \nab_\a
\psi^{(3,1)}_{\e} + \frac{1}{8} \nab_b h^{(3,1)}{}_{\a}{}^{\c}
(\c^b \c_a)_{\c}{}^{\a} .
\ee

The correction to $\nab_\a \hat{\L}$ is then given by

\be
(\nab_\a \hat{\L}_{b}{}^{\c})^{(3,1)} = -(\c_b)^{\c \d} \nab_\a
\psi^{(3,1)}_\d  +\frac{1}{4} (\nab_b \cF_{cd})^{(3,1)}
(\c^{cd})_{\a}{}^{\c} + \nab_b h^{(3,1)}{}_{\a}{}^{\c} 
\ee

This verifies that the corrections to the theory are indeed specified
just by fixing the deformation of the $\cF$-constraint which is given
by $J$. We have calculated the first possible deformation of the
constraint at order $l^4$ and cubic in fields and hence the first
derivative deformation of the Born-Infeld theory allowed by
supersymmetry.

One could also explicitly calculate the corrections to the
kappa-variations of the fields. These are of the same form as in the
undeformed case. For the variations of the coordinates, one simply has
to bear in mind that there are corrections to the embedding matrix
induced by $J$ (\ref{hJscalar},\ref{hJ2form},\ref {hJ4form}). For the
variation of the gauge field  one must account for the non-zero
components $\cF_{\a \b}$ and $\cF_{\a b}$. However, it should be
emphasised that a formulation where the symmetry is manifest is
preferable to one where explicit variations are required.

\section{Conclusions}

Using the superembedding formalism, we have shown that derivative corrections
to the D9-brane effective action can be systematically computed in a
manifestly kappa-symmetric manner. All supersymmetric and kappa-symmetric
deformations of Born-Infeld theory can be identified with elements of a
spinorial cohomology group. We have calculated explicitly the first such
possible correction at leading order in the dimensionful parameter $l$ and
leading order in number of fields. In general, one expects corrections
to the term we have found both at higher order in number of fields and
at higher order in $l$ in order to consistently solve the Bianchi
identity. Any such higher order completion will necessarily not be
unique due to the presence of higher order elements of the cohomology
group. Given that some all order completion should exist, the term we
have presented defines a kappa-symmetric theory which, upon gauge
fixing, must reproduce the leading derivative correction to the
four-point amplitude of the open superstring. It would be interesting
to find an explicit form for such an invariant to all orders, although
this would not be the complete effective action for the open
superstring because it would not contain (for example) higher
derivative four-point functions. It would also be of interest to find
the correct $J$ which would reproduce the full four-point interactions
to all orders in $l$. One would require some input from string theory
to fix the coefficients of the independent terms in such a
calculation, as discussed in \cite{dre03}. 
 
The constructions of \cite{hklt03} and the present work could be extended
to branes with both gauge and transverse bosonic degrees of
freedom. However, in the case of low (but not zero) codimension, there are two
constraints to deform, the embedding constraint and the
$\cF$-constraint. We expect that in such cases the solution to the
deformation problem is unique at order $l^4$. 

One can make use of our results in two ways: one could assume
string theory dualities, and thus construct the effective theories for
other D-branes by T-duality. Alternatively, one can perform the direct
calculation, using the same method to check if there is in fact only
one solution, thus guaranteeing that T-duality is
respected to order $l^4$ by supersymmetry.

\section*{Acknowledgments}
The authors would like to thank Paul Howe for useful discussions. SFK would
like to thank the German National Academic Foundation (Studienstiftung des
deutschen Volkes) for support.

\section*{Appendix}

We use the highest weight notation for irreducible representations. The 
relevant group is $Spin(1,9)$ which is D5 in the Cartan
classification. The relevant irreducible representations are given by:

\begin{align}
(10000) & \text{ vector } \notag \\
(01000) & \text{ two-form }\notag  \\
(11000) & \text{ traceless vector two-form } \notag \\
(00001) & \text{ downstairs spinor } \notag  \\
(00010) & \text{ upstairs spinor } \notag \\
(10010) & \text{ gamma-traceless (upstairs) vector spinor } \notag \\
(00002) & \text{ anti-self dual 5-form } \notag \\
(00003) & \text{ anti-self-dual 5-form (downstairs) spinor } \notag 
\end{align}

\subsection*{Scalars at $l^4$}
There are two ways of constructing the representation (00000) which
are cubic in the fields. They are,

\begin{align}
\nab \L \L \cF \text{  :  }
\phi_1 &= l^4\nab_b \L_{c}{}^{\a} \L^{b \b} \cF^{cd} (\c_d)_{\a \b}, \\
\L \L \nab \cF \text{  :  }
\phi_2 &= l^4\L_{a}{}^{\a} \L_{b}{}^{\b} \nab_c \cF^{ab} (\c^c)_{\a \b}. 
\end{align}

\subsection*{Field redefinitions at $l^4$}
There are fourteen ways of constructing the representation (00001)
which are cubic in the fields:

\begin{align}
\L \nab^2 \cF \cF \text {  :  }
(\d A_1)_\a &= l^4\L_{a}{}^{\b} \nab^a \nab^b \cF^{cd} \cF_{bc}
(\c_d)_{\a \b}, \\
(\d A_2)_\a &= l^4\L_{a}{}^{\b} \nab^a \nab^b \cF^{cd}
\cF_{b}{}^{e}(\c_{cde})_{\a \b}, \\
\nab \L \nab \cF \cF \text {  :  }
(\d A_3)_\a &= l^4\nab_a \L_{b}{}^{\b} \nab^a \cF^{bc} \cF_{cd}
(\c^d)_{\a \b}, \\
(\d A_4)_\a &= l^4\nab_a \L_{b}{}^{\b} \nab^a \cF_{cd} \cF^{bc}
(\c^d)_{\a \b}, \\
(\d A_5)_\a &= l^4\nab_a \L^{b \b} \nab_c \cF^{ad} \cF_{bd} (\c^c)_{\a
  \b}, \\
(\d A_6)_\a &= l^4\nab_a \L_{b}{}^{\b} \nab^a \cF^{bc} \cF^{de}
(\c_{cde})_{\a \b}, \\
(\d A_7)_\a &= l^4\nab_a \L_{b}{}^{\b} \nab^a \cF^{cd} \cF^{be}
(\c_{cde})_{\a \b}, \\
\L \nab \cF \nab \cF \text{  :  }
(\d A_8)_\a &= l^4\L_{a}{}^{\b} \nab^a \cF_{bc} \nab^b \cF^{cd}
(\c_d)_{\a \b}, \\
(\d A_9)_\a &= l^4\L_{a}{}^{\b} \nab_b \cF^{ac} \nab_c \cF^{bd}
(\c_d)_{\a \b}, \\
(\d A_{10})_\a &= l^4\L_{a}{}^{\b} \nab_b \cF^{ca} \nab_c \cF_{de}
(\c^{bde})_{\a \b}, \\ 
(\d A_{11})_\a &= l^4\L_{a}{}^{\b} \nab^a \cF_{bc} \nab^c \cF_{de}
(\c^{bde})_{\a \b}, \\
\nab \L \L \L \text{  :  }
(\d A_{12})_\a &= l^4\nab_a \L_{b}{}^{\b} \L^{a \c} \L^{c \d} (\c_c)_{\a
  \b} (\c^b)_{\c \d}, \\
(\d A_{13})_\a &= l^4\nab_a \L_{b}{}^{\b} \L^{a \c} \L^{b \d}
  (\c_{cde})_{\a \b} (\c^{cde})_{\c \d}, \\
(\d A_{14})_\a &= l^4\nab_a \L_{b}{}^{\b} \L^{a \c} \L^{c \d}
  (\c_{cde})_{\a \b} (\c^{bde})_{\c \d}.
\end{align}

Two linear combinations of these are exact in the sequence
(\ref{seq}). One finds, 

\begin{align}
\D_0 \phi_1 &= -\frac{1}{2} (\d A_1) - \frac{1}{4} (\d A_2)
+ \frac{1}{2} (\d A_4) + \frac{1}{4} (\d A_7) 
- \frac{i}{4} (\d A_{12}) + \frac{i}{8} (\d A_{14}) ,\\
\D_0 \phi_2 &=  (\d A_9) - \frac{1}{2} (\d A_{10}) + 2i (\d A_{12}) .
\end{align}

We can use these relations to eliminate $(\d A_{10})$ and $(\d
A_{14})$ when computing which combinations of $J$s are exact in the
sequence (\ref{seq}).

\subsection*{Deformations at $l^4$}
There are nineteen ways of constructing the representation (00002)
which are cubic in the fields. They are given below as tensors $J_{a_1
  a_2 a_3 a_4 a_5}$. It is understood
that one antisymmetrises over the free indices $a_1...a_5$ and explicitly
subtracts the dual.

\begin{align}
\nab^2 \L \L \cF \text{  :  }
J_1 &= l^4\nab_b \nab_c \L_{a_1}{}^{\a} \L^{b \b} \cF^{c}{}_{a_2}
(\c_{a_3 a_4 a_5})_{\a \b} , \\ 
\L \L \nab^2 \cF \text{  :  }
J_2 &= l^4\L_{b}{}^{\a} \L_{c}{}^{\b} \nab^b \nab^c \cF_{a_1 a_2}
(\c_{a_3 a_4 a_5})_{\a \b} , \\
J_3 &= l^4\L_{b}{}^{\a} \L_{a_1}{}^{\b} \nab^b \nab^c \cF_{a_2 a_3}
(\c_{a_4 a_5 c})_{\a \b} , \\
J_4 &= l^4\L_{b}{}^{\a} \L_{c}{}^{\b} \nab^d \nab_{a_1} \cF^{bc} (\c_{a_2
  a_3 a_4 a_5 d})_{\a \b} , \\
\nab \L \nab \L \cF \text{  :  }
J_5 &= l^4\nab_b \L_{a_1}{}^{\a} \nab^b \L_{a_2}{}^{\b} \cF_{a_3 a_4}
(\c_{a_5})_{\a \b} , \\
J_6 &= l^4\nab_b \L_{c}{}^{\a} \nab^b \L^{c \b} \cF_{a_1 a_2}
(\c_{a_3 a_4 a_5})_{\a \b} , \\
J_7 &= l^4\nab_b \L_{c}{}^{\a} \nab^b \L_{a_1}{}^{\b} \cF^{c}{}_{a_2}
(\c_{a_3 a_4 a_5})_{\a \b} , \\
J_8 &= l^4\nab_b \L_{a_1}{}^{\a} \nab_c \L_{a_2}{}^{\b} \cF^{bc} (\c_{a_3
  a_4 a_5})_{\a \b} , \\
\nab \L \L \nab \cF \text{  :  }
J_9 &= l^4\nab_b \L_{a_1}{}^{\a} \L_{a_2}{}^{\b} \nab^b \cF_{a_3 a_4}
(\c_{a_5})_{\a \b} , \\
J_{10} &= l^4\nab_b \L_{c}{}^{\a} \L^{b \b} \nab^c \cF_{a_1 a_2} (\c_{a_3
  a_4 a_5})_{\a \b} , \\
J_{11} &= l^4\nab_b \L_{c}{}^{\a} \L_{a_1}{}^{\b} \nab^b \cF^{c}{}_{a_2}
(\c_{a_3 a_4 a_5})_{\a \b} , \\
J_{12} &= l^4\nab_b \L_{a_1}{}^{\a} \L_{a_2}{}^{\b} \nab^b
\cF^{c}{}_{a_3} (\c_{a_4 a_5 c})_{\a \b} , \\
J_{13} &= l^4\nab_b \L_{a_1}{}^{\a} \L_{c}{}^{\b} \nab^b \cF^{c}{}_{a_2}
(\c_{a_3 a_4 a_5})_{\a \b} , \\
J_{14} &= l^4\nab_b \L_{a_1}{}^{\a} \L_{c}{}^{\b} \nab_{a_2} \cF^{bc}
(\c_{a_3 a_4 a_5})_{\a \b} , \\
J_{15} &= l^4\nab_b \L_{a_1}{}^{\a} \L^{b \b} \nab^c \cF_{a_2 a_3}
(\c_{a_4 a_5 c})_{\a \b} , \\
J_{16} &= l^4\nab_b \L_{c}{}^{\a} \L^{b \b} \nab^d \cF^{c}{}_{a_1}
(\c_{a_2 a_3 a_4 a_5 d})_{\a \b} , \\
\nab^2 \cF \nab \cF \cF \text{  :  }
J_{17} &= l^4\nab_b \nab_c \cF_{a_1 a_2} \nab^b \cF^{c}{}_{a_3} \cF_{a_4
  a_5} , \\
J_{18} &= l^4\nab_b \nab_c \cF_{a_1 a_2} \nab^b \cF_{a_3 a_4}
\cF^{c}{}_{a_5} , \\
\nab \cF \nab \cF \nab \cF \text{  :  }
J_{19} &= l^4\nab_b \cF_{c a_1} \nab^b \cF_{a_2 a_3} \nab^c \cF_{a_4 a_5}. 
\end{align}

From the twelve independent spinors given previously we derive the
following twelve exact combinations of the above $J$s which correspond
to field redefinitions and not genuine deformations,

\begin{align}
E_1 = \D_1 (\d A_1) &= 16 J_1 + 4 J_2 + J_4 , \\
E_2 = \D_1 (\d A_2) &= 96 J_{18} + i [16 J_1 + 4 J_2 - 48 J_3 + J_4]
, \\
E_3 = \D_1 (\d A_3) &= J_6 - 2 J_7 + 2 J_{11} , \\
E_4 = \D_1 (\d A_4) &= 2 J_7 + J_{10} - 2 J_{13} , \\
E_5 = \D_1 (\d A_5) &= 8 J_7 + 8 J_8 - 2 J_{10} + 8 J_{14} + J_{16}
, \\
E_6 = \D_1 (\d A_6) &= -96 J_{17} + i [...] , \\
E_7 = \D_1 (\d A_7) &= 6 J_{18} + i [J_7 - 6 J_9 - J_{11} + 3
  J_{12}] , \\
E_8 = \D_1 (\d A_8) &= 2 J_{10} + 8 J_{13} + 8 J_{14} + J_{16} , \\
E_9 = \D_1 (\d A_9) &= 2 J_{10} - 8 J_{11} - 8 J_{14} - J_{16} , \\
E_{10} = \D_1 (\d A_{11}) &= J_{19} + i [...]     , \\
E_{11} = \D_1 (\d A_{12}) &= 6 J_9 - J_{11} + 3 J_{12} + J_{13} - 3
J_{15} , \\ 
E_{12} = \D_1 (\d A_{13}) &= J_2 - J_{10}  .
\end{align}

The notation $[...]$ is used above to denote a linear combination of
terms involving $\L$ (the terms $J_1$ to $J_{16}$). The details of the
linear combination are not needed for the ensuing analysis. 
The above relations mean that it is consistent to remove all but $J_3
, J_4 , J_5 , J_9 , J_{10} , J_{14} , J_{16}$ from the set of $J$s to
check which satisfy closure non-trivially.

\subsection*{Closure constraints at $l^4$}
There are six ways of constructing the representation (00003) which
are cubic in the fields. They are given below as tensors, $C^{a_1 a_2
  a_3 a_4 a_5}{}_{\d}$. It is understood that
one should antisymmetrise the free indices $a_1...a_5$, explicitly
subtract the dual and take the $\c$-traceless part.

\bea
\nab \L \nab \L \L \text{  :  }
C_1 &=& l^4\nab_b \L^{a_1 \a} \nab_c \L^{a_2 \b} \L^{c \c}
(\c^{a_3 a_4 a_5})_{\a \b} (\c^b)_{\c \d} \\ 
C_2 &=& l^4\nab_b \L_{c}{}^{\a} \nab^c \L^{a_1 \b} \L^{a_2 \c}
(\c^{a_3 a_4 a_5})_{\a \b} (\c^b)_{\c \d} \\ 
\L \L \nab^2 \L \text{  :  }
C_3 &=& l^4\L_{b}{}^{\a} \L^{a_1 \b} \nab^b \nab_c \L^{a_2 \c}
(\c^{cd a_3 a_4 a_5})_{\a \b} (\c_d)_{\c \d} \\ 
\nab \L \nab^2 \cF \cF \text{  :  }
C_4 &=& l^4\nab_b \L^{a_1 \a} \nab^b \nab^c \cF^{a_2 a_3} \cF^{a_4 a_5}
(\c_c)_{\a \d} \\ 
\nab \L \nab \cF \nab \cF \text{  :  }
C_5 &=& l^4\nab_b \L^{a_1 \a} \nab^b \cF^{a_2 a_3} \nab^c \cF^{a_4 a_5}
(\c_c)_{\a \d} \\
\L \nab^2 \cF \nab \cF \text{  :  }
C_6 &=& l^4\L^{a_1 \a} \nab_b \nab_c \cF^{a_2 a_3} \nab^b \cF^{a_4 a_5}
(\c^c)_{\a \d} . 
\eea

Applying $\D_2$ to the nineteen $J$s we find:

\begin{center}
\be
\begin{array}{lll}
\D_2 J_1 = \frac{i}{2} C_3 & \D_2 J_8 = -2iC_1 & \D_2
J_{15} = -2 C_5 \\ 
\D_2 J_2 = 0 & \D_2 J_9 = -\frac{1}{2}(C_5 +C_6) \hspace{20pt} &  \D_2
J_{16} = -8iC_1 \\ 
\D_2 J_3 = -2C_6 & \D_2 J_{10} = 0 & \D_2 J_{17} = -iC_4 \\
\D_2 J_4 = -8iC_3 \hspace{20pt} & \D_2 J_{11} = -iC_2 & \D_2
J_{18} = -iC_6\\ 
\D_2 J_5 = C_4 & \D_2 J_{12} = C_6 -C_5 & \D_2 J_{19} = -iC_5 \\
\D_2 J_6 = 0 & \D_2 J_{13} = -iC_2 & \\
\D_2 J_7 = -iC_2 & \D_2 J_{14} = i(C_1 +C_2) &
\end{array}
\ee
\end{center}

Thus we can see that from the set, $\{ J_3 , J_4 , J_5 , J_9 , J_{10} ,
J_{14} , J_{16} \} $, the only closed $J$ is given by $J_{10}$. This
is the fist non-trivial deformation of the $\cF$-constraint.

In constructing the terms in the preceding sections the program LiE,
\cite{lie} proved useful. We used it to check that we found the
correct number of terms of each representation. We have also checked some
of the gamma-matrix manipulations with the Mathematica package GAMMA 
\cite{gamma}.

\end{document}